\documentclass{jfm}
\usepackage{graphicx}
\usepackage{natbib}
\usepackage[T1]{fontenc}
\usepackage[latin9]{inputenc}
\usepackage{color}
\usepackage{textcomp}
\usepackage{amsmath}
\usepackage{floatflt}
\usepackage{float}
\usepackage{wrapfig}
\usepackage{capt-of}
    
\ifCUPmtlplainloaded \else
  \checkfont{eurm10}
  \iffontfound
    \IfFileExists{upmath.sty}
      {\typeout{^^JFound AMS Euler Roman fonts on the system,
                   using the 'upmath' package.^^J}%
       \usepackage{upmath}}
      {\typeout{^^JFound AMS Euler Roman fonts on the system, but you
                   dont seem to have the}%
       \typeout{'upmath' package installed. JFM.cls can take advantage
                 of these fonts,^^Jif you use 'upmath' package.^^J}%
      }
  \else
  \fi
\fi

\ifCUPmtlplainloaded \else
  \checkfont{msam10}
  \iffontfound
    \IfFileExists{amssymb.sty}
      {\typeout{^^JFound AMS Symbol fonts on the system, using the
                'amssymb' package.^^J}%
       \usepackage{amssymb}%

      }{}
  \fi
\fi

\ifCUPmtlplainloaded \else
  \IfFileExists{amsbsy.sty}
    {\typeout{^^JFound the 'amsbsy' package on the system, using it.^^J}%
     \usepackage{amsbsy}}
    {}
\fi

\makeatletter

\DeclareTextCommandDefault{\regimeN}{\textcircled{%
      \check@mathfonts\fontsize\sf@size\z@\math@fontsfalse\selectfont N}}
\DeclareTextCommandDefault{\regimeR}{\textcircled{%
      \check@mathfonts\fontsize\sf@size\z@\math@fontsfalse\selectfont R}}
\DeclareTextCommandDefault{\regimeC}{\textcircled{%
      \check@mathfonts\fontsize\sf@size\z@\math@fontsfalse\selectfont C}}
\DeclareTextCommandDefault{\regimeG}{\textcircled{%
      \check@mathfonts\fontsize\sf@size\z@\math@fontsfalse\selectfont G}}
\DeclareTextCommandDefault{\regimeU}{\textcircled{%
      \check@mathfonts\fontsize\sf@size\z@\math@fontsfalse\selectfont U}}
\DeclareTextCommandDefault{\regimeA}{\textcircled{%
      \check@mathfonts\fontsize\sf@size\z@\math@fontsfalse\selectfont A}}
\DeclareTextCommandDefault{\regimeS}{\textcircled{%
      \check@mathfonts\fontsize\sf@size\z@\math@fontsfalse\selectfont S}}
\DeclareTextCommandDefault{\regimeB}{\textcircled{%
      \check@mathfonts\fontsize\sf@size\z@\math@fontsfalse\selectfont B}}

\makeatother

\title{New patterns in high-speed granular flows}

\author[N. Brodu, R. Delannay, A. Valance and P. Richard]%
{Nicolas Brodu$^1$%
  \thanks{Email address for correspondence: nicolas@brodu.net},\ns
Renaud Delannay$^1$\break
Alexandre Valance$^1$\break
and Patrick Richard$^2$}

\affiliation{$^1$Institut de Physique de Rennes, UMR CNRS 6251, Universit\'e de Rennes
1, Campus de Beaulieu B\^atiment 11A, 263 av. G\'en\'eral Leclerc,
35042 Rennes CEDEX, France\\[\affilskip]
$^2$LUNAM Universit\'e, IFSTTAR, GPEM, site de Nantes, Route de Bouaye, 44344 Bouguenais cedex, France}

\pubyear{2014}
\volume{?}
\pagerange{?--?}
\date{?; revised ?; accepted ?}

\begin{document}

\maketitle

\begin{abstract}
We report on new patterns in high-speed flows of granular materials obtained
by means of extensive numerical simulations. These patterns emerge from the destabilization
of unidirectional flows upon increase of mass holdup and inclination
angle, and are characterized by complex internal structures
including secondary flows, heterogeneous particle volume fraction, symmetry breaking and
dynamically maintained order. In particular, we evidenced steady and fully developed "supported"
flows, which consist of a dense core surrounded
by a highly energetic granular gas. Interestingly, despite their overall diversity,
these regimes are shown to obey a scaling law
for the mass flow rate as a function of the mass holdup.
This unique set of 3D flow regimes raises new challenges
for extending the scope of current granular rheological models.
\end{abstract}


\section{Introduction}
Granular gravity-driven flows are very common in industrial and geophysical processes. 
These flows are generally dense and can be confined by lateral walls or levees (due to self-channeling). 
The scientific community has paid particular attention to these flows 
over the last thirty years.  
However, their modeling is still an open issue.
The complexity comes from grain/grain interactions that include both collisions and long lasting frictional contacts. 
Identifying regions of the flow where one type of interaction 
prevails over the other is part of the issue to be resolved.

One of the most studied configurations is the inclined plane geometry.
Partly because it is a simple and good model for many common situations, 
but also because it may be seen as a rheological test with constant friction. Indeed, if sidewall friction
is negligible, 
for steady and fully developed (SFD) flows, 
the tangential and normal forces on the base correspond exactly to the components of the flow weight. 
Their ratio, which is nothing but the apparent friction $\mu$, is equal to the tangent of the angle of inclination $\theta$.
To date, experiments and simulations have focused mainly on flows with moderate inclination, leading to fairly simple unidirectional 
SFD flows \citep{GDRMIDI2004,Delannay2007}. 
However, more complex SFD flows with spanwise vortices were obtained 
for higher angles \citep{Borzsonyi_etal_2009}. 
One therefore expects that upon further increase of the inclination angle
more and more complex flow features should emerge.

In the case of flows running on a flat frictional base,
the ratio of the tangential to the normal component of the contact force acting
on a grain in contact with the base has an upper bound which is the microscopic friction coefficient $\mu_m$.  
Thereby, the effective friction $\mu=\tan \theta$ is also bounded by $\mu_m$. For definite and 
realistic values of $\mu_m$, this automatically 
limits the possible angles for SFD flows. In the case of a bumpy base, there is also a limit, which depends in a complex and intricated manner on the microscopic friction coefficient, the bumpiness and the coefficient of restitution $e$ \citep{GDRMIDI2004}. In particular, a small restitution coefficient allows SFD flows at high angles of incline. However, these flows seem to be unstable \citep{Taberlet2007}, which makes them difficult to study. The easiest way to obtain SFD flows at high angles is to introduce frictional side walls. This is what we have done in the present work.
If the grain/wall friction coefficient is high enough, one may expect that the base 
friction supplemented by the side wall friction will be able to balance the driving 
component of the weight.

We have conducted simulations of granular flows down flat and steep inclines with frictional side walls using a discrete element model (DEM).
The principle of DEM simulations is to treat each grain as a sphere (of diameter $D$) subject to gravity and contact forces with both the other grains and the basal and lateral walls. These contact forces are characterized by a coefficient of restitution $e_{g}=0.972$, a coefficient of friction $\mu_{g}=0.33$ for interactions between grains, with $e_{m}=0.8$ and $\mu_{m}=0.596$  for grain/wall interactions (values taken from \citep{Louge_Keast_2001}). Newton's second law is applied to calculate the motion of each individual particle. This method, which has successfully been used to simulate granular flows \citep{Cundall_Strack_1979,Silbert_etal_2001,Luding2008},
has been optimized to obtain three-dimensional SFD flows within a reasonable computation time. 
\begin{floatingfigure}[r]{0.5\columnwidth}
\includegraphics[width=0.45\columnwidth]{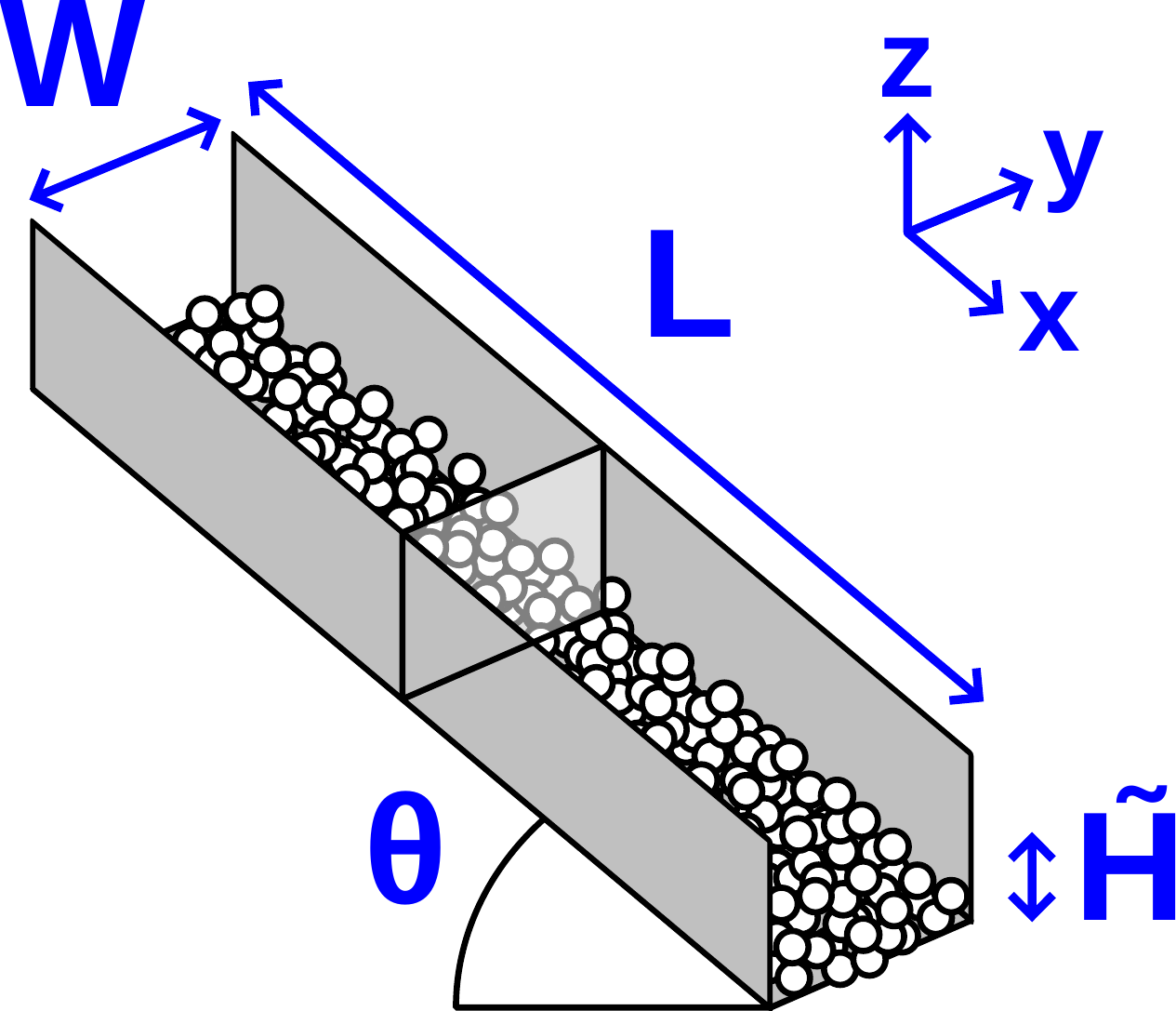} 
\caption{Geometry of the system. The grains flow on an inclined plane (length $L$) and between two side walls separated by a gap $W$. Periodic boundary conditions are used in the $x$-direction. The variations of the mass holdup $\tilde{H}$ and of the angle of inclination $\theta$ lead to the observation of new SFD flow regimes.}
\label{fig:geometry}
\end{floatingfigure}
We used periodic boundary conditions in the streamwise direction, while walls constrain the flow in lateral directions (see Fig.~\ref{fig:geometry}). The periodic cell had a length $L=20D$ and width $W=68D$. We used a relative large cell width to allow for the development of transverse instabilities

\noindent\citep{Borzsonyi_etal_2009}. It is worth mentioning that most simulations in the literature used lateral periodic boundary conditions with a small width (typically $W=20D$), preventing three-dimensional patterns to develop.

The mass hold-up is a measure of the mass of particles per unit basal surface within the simulation cell: $\tilde{m}=\sum_g m_g/A$, where $m_g$ is the mass of a grain $g$ and $A=L\times W$ is the basal area of the cell. For ease of interpretation, we choose to express the mass hold-up in terms of an equivalent grain height $\tilde{H}$ by dividing $\tilde{m}$ by the particle density $\rho$: $\tilde{H}=\tilde{m}/(\rho D)$. $\tilde{H}$ is thus the height of a dense block with the same volume as the grains. This quantity is a control parameter which simply specifies the amount of particles within the system, irrespectively of their spatial repartition. For each value of the control parameters $\theta$ and $\tilde{H}$, the simulations were run up to a stabilization of the total kinetic energy of the system, see \citep{Brodu_etal_2013} for details as we reused the same configuration and simulation parameters.
We obtained SFD flows for all the flow configurations we have investigated, varying extensively the inclination angle $\theta$ between $0$ and $50\text{\textdegree}$ and the mass hold-up $\tilde{H}$ between $0$ and $20$. We also tested much larger values of the angle of inclination (see Supplementary Figure 2), but we always reached SFD flows after a transient which duration increases with the angle.

\begin{figure*}
\begin{center}
\includegraphics[width=1.0\columnwidth]{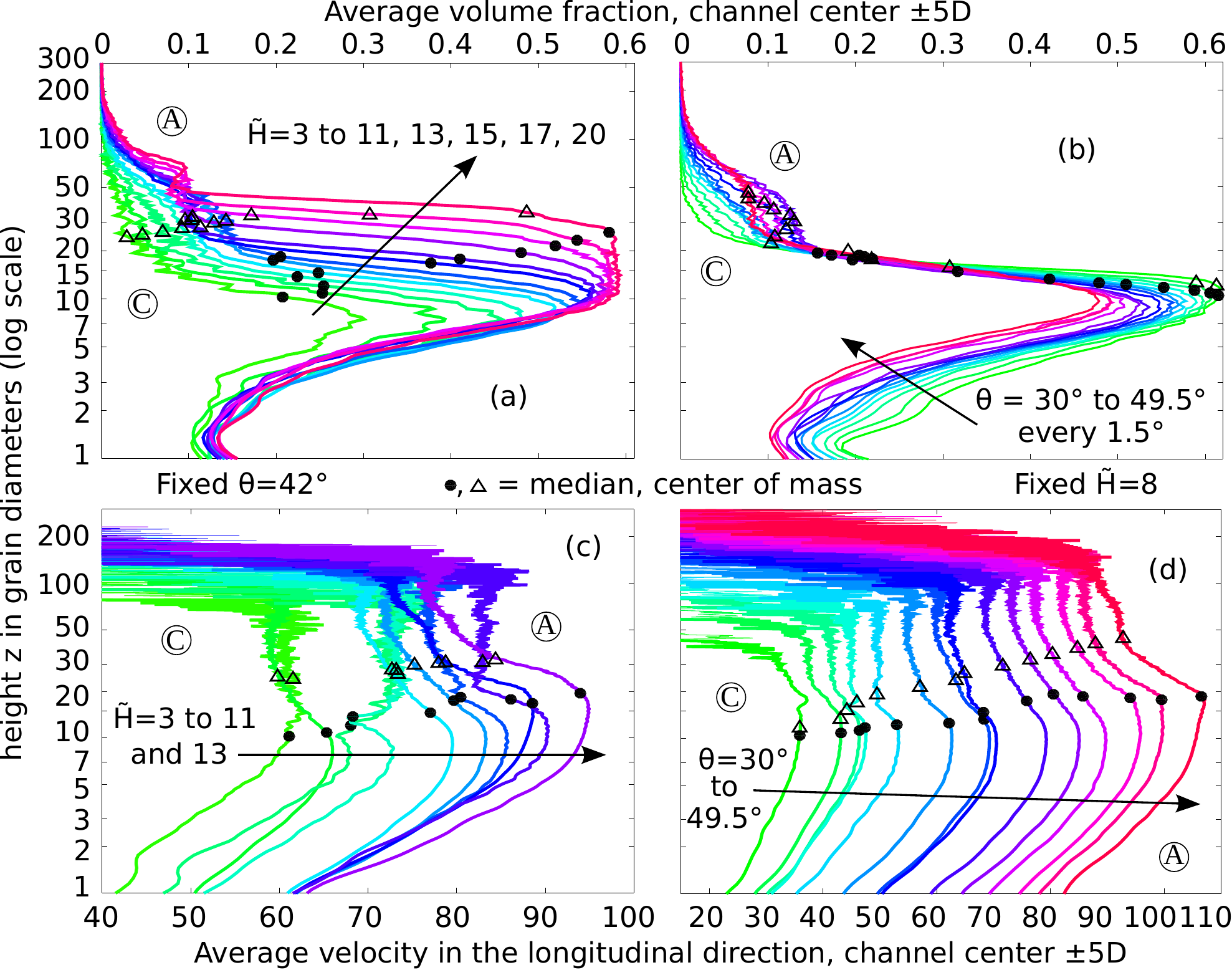}
\end{center}
\caption{
Vertical profiles of the volume
fraction of the flow ((a) and (b)), of the velocity in the main flow direction X ((c) and (d)).
All quantities are measured at the center of the channel and are averaged over $10D$ in the transverse direction Y.
The curves reported in the right column are for a fixed mass holdup
$\tilde{H}=8$ and for different angles of inclination.
Those reported in the left are for a fixed angle of inclination
$\theta=42\text{\textdegree}$ and for different values
of mass holdup. 
Circles and triangles respectively indicate the vertical position
of the median and center of mass, which increases both with the mass holdup and the angle of inclination.}
\label{fig:cprofiles}
\end{figure*}

Our simulations reveal the existence of many SFD regimes which have very different characteristics: coexistence of order and disorder phases, symmetry breaking, oscillations, intermittency, stacked granular "convection" rolls, polyphasic flows, etc. These new regimes emerge from the destabilization of SFD unidirectional flows upon increase of the mass holdup and the slope. In a previous work \citep{Brodu_etal_2013} flows corresponding to $\tilde{H}=4$ and $\theta<23^\circ$ were studied. We thus focus more specifically here on large inclination flows ($\theta>30^\circ$) and on the effect produced by an increasing mass holdup. The next section gives details on supported regimes, observed for $\theta>30^\circ$. Section 3 gives an overview of the different regimes we obtained by exploring the parameter space, and reports their domain of existence. Section 4 reveals that, despite the very different characteristics of the observed regimes, they show common features. Concluding remarks are given in section 5.\\

\section{Supported regimes\label{sec:supported_regimes}}

We identified at high angle ($\theta>30^\circ$) a new flow regime, referred to as "supported flows". This regime is drastically different from those reported in the literature and, in particular, from the "rolls" regime investigated in several recents works \citep{Forterre_Pouliquen_2002,Borzsonyi_etal_2009,Brodu_etal_2013}. Dense unidirectional flows destabilize upon increase of the inclination angle (typically between $20^\circ$ and $30^\circ$) and then exhibit longitudinal rolls. The above references provide evidence that, with these rolls, the particle volume fraction becomes lower at the base than in the midst of the flow. Density inverted profiles are also predicted by the granular kinetic theory \citep{Jenkins_Askari_1999}. However, in these regimes with rolls, the depletion of particles is moderate and located at the flow base. The geometric structure remains that of a slab of grains occupying the whole width of the channel, together with the associated secondary circulation pattern. Upon further increase of the angle, $\theta>30^\circ$, a strongly sheared, dilute and agitated layer spontaneously appears at the base of the flow (see Fig.~\ref{fig:Temperature}).
\begin{floatingfigure}[r]{0.5\columnwidth}
\includegraphics[width=0.45\columnwidth]{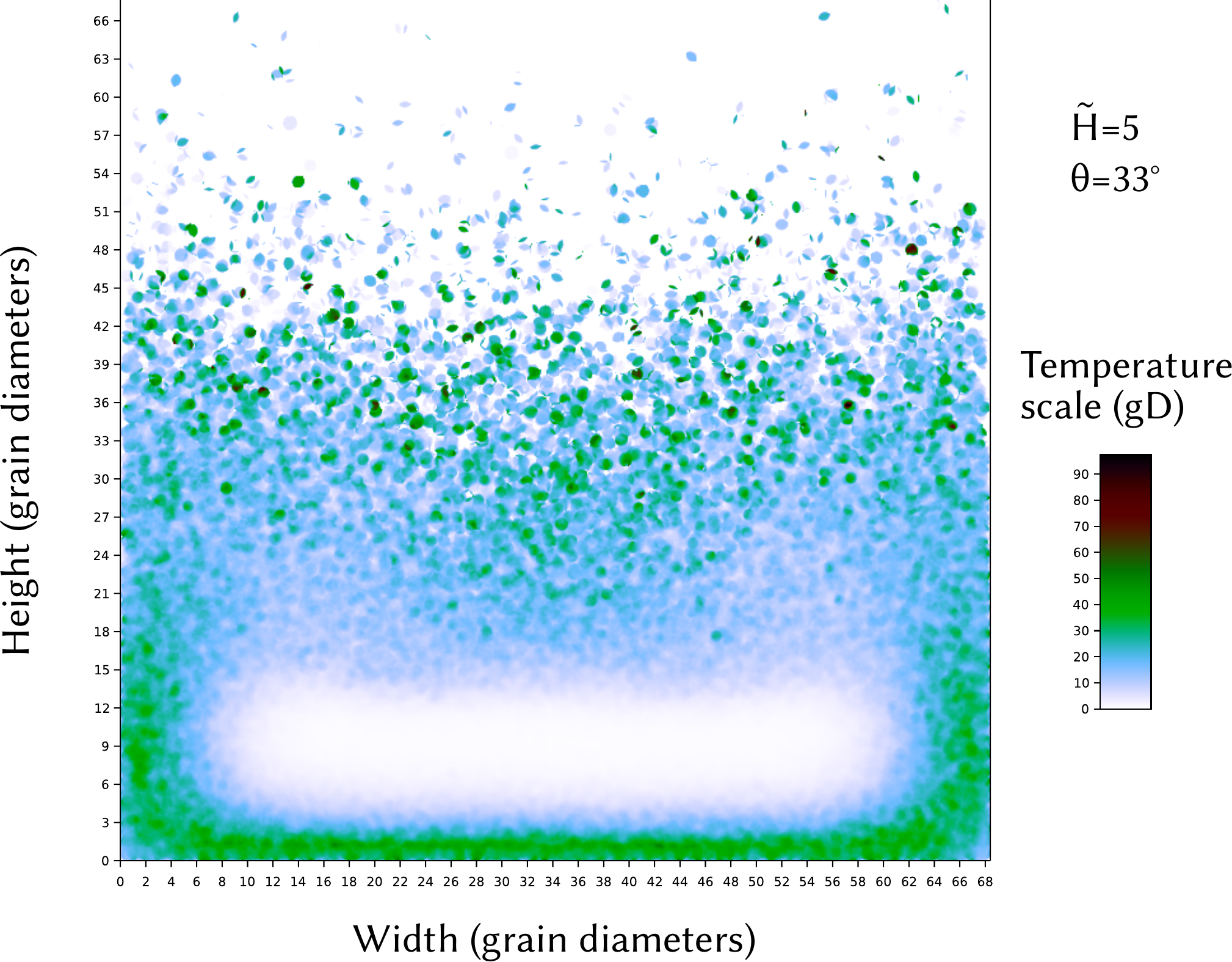}
\caption{Map of the granular "temperature" for $\tilde{H}=5$ and $\theta=33^\circ$, the same situation as for the bottom-right snapshot in Fig. 5. This clearly shows a much less agitated dense core than the surrounding "hot" basal and side layers of gas.}
\label{fig:Temperature}
\end{floatingfigure}
Such a layer is able to support a dense packing of grains moving as a whole. Additional granular gas layers separate that dense core from the side walls. This geometric structure is clearly distinct from that of the dense slab with secondary rolls, as can be seen on the bottom line of snapshots of Fig.~\ref{fig:PhaseDiagram} (see also Supplementary Figure 2), which also shows their transition. In particular, the packing fraction in the core of the supported regimes is higher than within the dense slab with the secondary rolls. This high unexpected volume fraction is not seen in density inverted profiles obtained from the kinetic theory of granular gas \citep{Jenkins_Askari_1999}. These supported regimes have been already mentioned in the literature
as a possible explanation for the unexpected high mobility of granular avalanches.
\cite{Campbell_1989} indeed suggested that the existence of a layer
of highly agitated particles at low concentration beneath a densely
packed main body could reduce the apparent basal friction and allow the
flow to reach long runouts. 
However, DEM simulations were unable up to now to reproduce
these flow regimes as steady and stable states. They were only observed 
as transient states in decelerated flows \citep{Campbell_1989} or as
a steady but unstable state at a unique value of the inclination angle \citep{Taberlet2007}.
In contrast, the supported regimes reported here are steady and fully developed, 
stable and, thanks to sidewall friction, were obtained within a large range of
inclination angles.

The existence of a stabilized dense core within a very agitated and dilute region
is probably a direct consequence of the clustering instability occurring
in granular gas \citep{McNamara_Young_1994}. It is also worth noting
that this flow regime bears a strong resemblance with that observed
experimentally by Holyoake and McElwaine \citep{Holyoake_McElwaine_2012}
on steep slopes with a "depletion layer" at the walls. The structure of supported flows may also bear some resemblance to vibrated beds \citep{Eshuis2013}. Gravity-driven flows can be seen as the superposition of two sollicitations: A shearing, induced by gravity; and collisions with the boundaries, which can be seen as either energy sources or sinks depending on their nature (e.g. loose base comprised of other mobile grains). In our situation, the base behaves as an energy source, which resembles the situation for vibrated media. Motion along the flow, mainly driven by gravity, could be more or less independent from motion in the transverse direction which, according to this hypothesis, would be mainly driven by the interactions of the grains with the boundaries. This would explain a similarity with some regimes observed in vibrated media. However, even if similarities may occur, a major difference is that in our case there is a volumic source of momentum, which induces shearing, while in the vibrated case, momentum is only transferred to the granular medium through the boundaries. It would nevertheless be interesting to use a hydrodynamic-like model \citep{Eshuis2013} to study the stability of the usual unidirectional flow when the control parameters increase, and the patterns which could result from the destabilisation.  

Fig.~\ref{fig:cprofiles} shows typical volume fraction and velocity profiles
for SFD supported regimes in the vertical ($x$,$z$) symmetry plane. They present a dense core, moving at a fast and almost uniform speed,
floating above a highly agitated granular gaseous phase, and toped by a dilute "atmosphere" which spans over a great distance over the dense core. Due to this heterogeneous mass distribution, the center of mass is located just on top of the core. The median of mass is unaffected by the distance of small contributions high in the dilute "atmosphere". By definition, the median separates half of the mass beneath and half above, so it arguably better corresponds to the intuition for where is the "middle" of the flow. We plot both quantities in Fig.~\ref{fig:cprofiles}.
When the mass holdup increases the core lifts up and densifies (see Fig.~\ref{fig:cprofiles}a). 
Its lateral width decreases with increasing $\tilde{H}$ because the lateral pressure pushes the grains toward
the central core (see Fig.~\ref{fig:PhaseDiagram}). This core can reach very high values of the volume
fraction up to $0.6$ at large mass holdup, while the volume fraction
in the supporting basal gaseous layer is below $0.2$. 

Above the dense core, the volume fraction is well described by
a decreasing exponential: $\nu(z)\propto\exp(-z/H_{C})$, where $H_{C}$
represents the characteristic height of the atmosphere (see Supplementary Figure~1).
The core slowly "evaporates" as the angle increases for a gradual
transition to granular gas at larger angles (see Supplementary Figures~2 and~3).
Surprisingly, when the angle $\theta$ increases, the
altitude of the core remains nearly constant (see Fig.~\ref{fig:cprofiles}b).
However, the center of mass of the flow lifts up and the core thickness decreases
as a non-negligible part of the material is transferred into the top
granular gaseous phase. The vertical expansion of the flow is necessary
to increase its friction on the lateral boundaries and to balance
the driving force, which increases with the inclination angle. The
basal friction cannot exceed $\mu_{m}Mg\cos\theta$, where $\mu_{m}=0.596$
is the microscopic friction value used in the simulations \citep{Brodu_etal_2013}
and $M$ is the mass of the grains. Thus, for large angles, a large
part of the friction comes from the lateral walls as discussed in more
details below. 

\begin{figure}[!b]
\begin{centering}
\includegraphics[width=1.0\columnwidth]{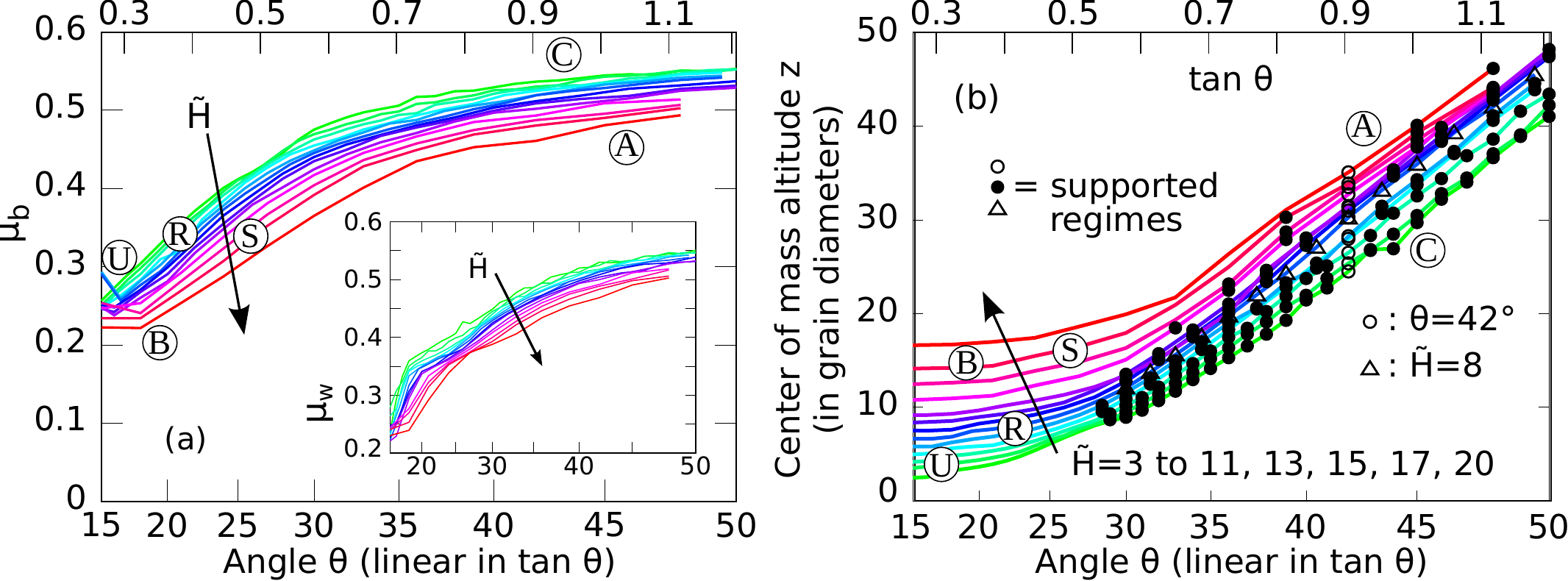} 
\par\end{centering}
\caption{
(a) Effective friction coefficients on the base $\mu_{b}$ and the wall $\mu_{w}$ (inset) as a function of
the inclination angle $\theta$. Both coefficients increase with $\theta$ but
show a reduction for increasing  mass hold-up.
(b) Altitude of the center of mass $C_{M}$ as a function of $\theta$ for various mass hold-up. 
Markers indicate the states corresponding to the supported flow regimes
for which $C_M$ exhibits a linear increase with $\tan\theta$.
Labels \regimeU,~\regimeR,~\regimeC,~\regimeA,~\regimeS~and \regimeB~refer to the
different flow regimes defined in Fig.~\ref{fig:PhaseDiagram}.
}
\label{fig:cmass_mubw}
\end{figure}

The effective friction coefficients respectively at the
base $\mu_{b}$ and at the walls $\mu_{w}$ are computed
as the ratio of tangential to normal stresses.
Fig.~\ref{fig:cmass_mubw}a shows the dependency of $\mu_{b}$
and $\mu_w$ on $\theta$ and $\tilde{H}$. 
Both coefficients increase and saturate at high inclination angles as they are upper-bounded by $\mu_{m}$.
In contrast, they decrease with the mass holdup: for a given angle,
the basal friction reduces as more matter is added in the flow. 
This reduction of basal friction with increasing mass hold-up
has never been reported before and may be a clue for explaining
the long run-out for large rock avalanches \citep{Campbell_1989}.
\vspace{1ex}

\hspace*{-\parindent}\begin{minipage}[t]{0.5\textwidth}
\hspace{2ex}
In the vertical direction however, friction on the walls do not significantly contribute to supporting the weight of the grains. As the center of mass lifts up (Figs. \ref{fig:cprofiles} and \ref{fig:cmass_mubw}b), the normal component of the integrated stress on the lateral walls is increased. The zone below the core then reaches a high pressure as it must sustain the flow weight. We have computed the ratio between the vertical component of the force exerted by the grains on the walls, to the vertical component of the force exerted by the grains on the base. This ratio is below 1.3\% in the non-supported regimes, with negative values in the supported ones (see Fig.~\ref{fig:wallcontrib}). Indeed, in these cases, grains escaping the high pressure gaseous region below the core rebound upward on the wall. Grains circulate slowly downward within the dense core to compensate. The very weak Janssen effect that is observed in the non-supported regimes is thus negated by the convection in the supported ones.
\end{minipage}
\hspace{0.02\textwidth}\begin{minipage}[t]{0.48\textwidth}
\raisebox{-\height}{\includegraphics[width=1\columnwidth]{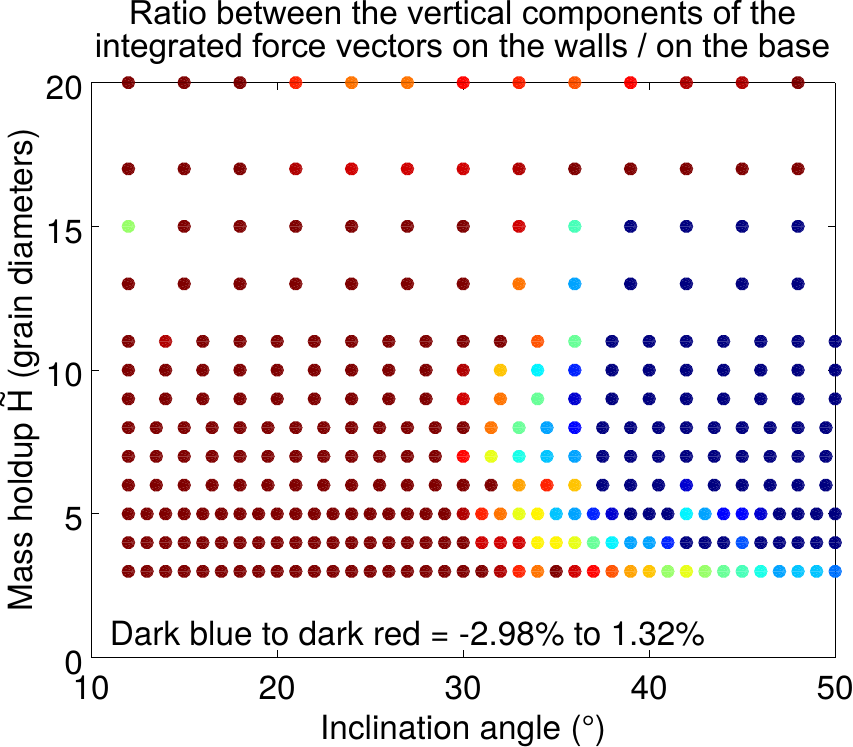}}
\captionof{figure}{\label{fig:wallcontrib}
Proportion of the effective weight retained by side walls. Positive values of the ratio are when the walls push up the grains, or equivalently when the grains push down on the side walls, since the base always pushes up. This is the case for the non-supported regimes, a Janssen effect that disappears in the supported regimes.}
\end{minipage}
\\

At the same time, the presence of a granular gas at the base lowers the friction compared to the dense regimes, hence larger velocities are reached in steady state. These large velocities are partially due to large sliding velocities and, for the rest, to large gradients, at the base (see Fig.~\ref{fig:cprofiles}c and ~\ref{fig:cprofiles}d). The top layer of gas only contributes marginally to this picture as it does not go faster than the core. However, as mentioned earlier, the vertical extent of the flow is a key feature to understand the balance between the
gravitational driving force and side walls friction. The position 
$C_{M}$ of the center of mass of the flow is a simple
and interesting indicator which is shown in   
Fig.~\ref{fig:cmass_mubw}b.  For the supported flows
(indicated by dots in the Figure), $C_{M}$ 
increases linearly with $\tan\theta$: $C_{M}=a\tan\theta+b$,
where the slope $a=54.6D$ is independent of the mass holdup. Using
a simple force balance, it can be shown that the slope is simply
given by $a\approx W/2\mu_m$ (see Appendix). In contrast,
the parameter $b$ increases with mass holdup and reflects
the corresponding increase of the core thickness with $\tilde{H}$
described in Fig. \ref{fig:cprofiles}a.

\section{Phase Diagram} 
In addition to the supported flows, we have discovered other new regimes
by exploring extensively and systematically the parameter space $(\theta,\tilde{H})$.
We report in Fig.~\ref{fig:PhaseDiagram} the domain of existence of the different regimes. These were identified using a combination of the flow structure (i.e., the concentration and the velocity fields) together with the evolution of the kinetic energy over time. We have decided to use large grayed-out transition bands for intermediate situations, together with dashed lines for transitions that appear sharp at our resolution, but which may not be so at a lower resolution. Some transitions are also visible in Fig.~\ref{fig:wallcontrib}. The regime identifiers are labeled by circled letters and are briefly described below:
\begin{figure*}
\begin{centering}
\includegraphics[width=\textwidth]{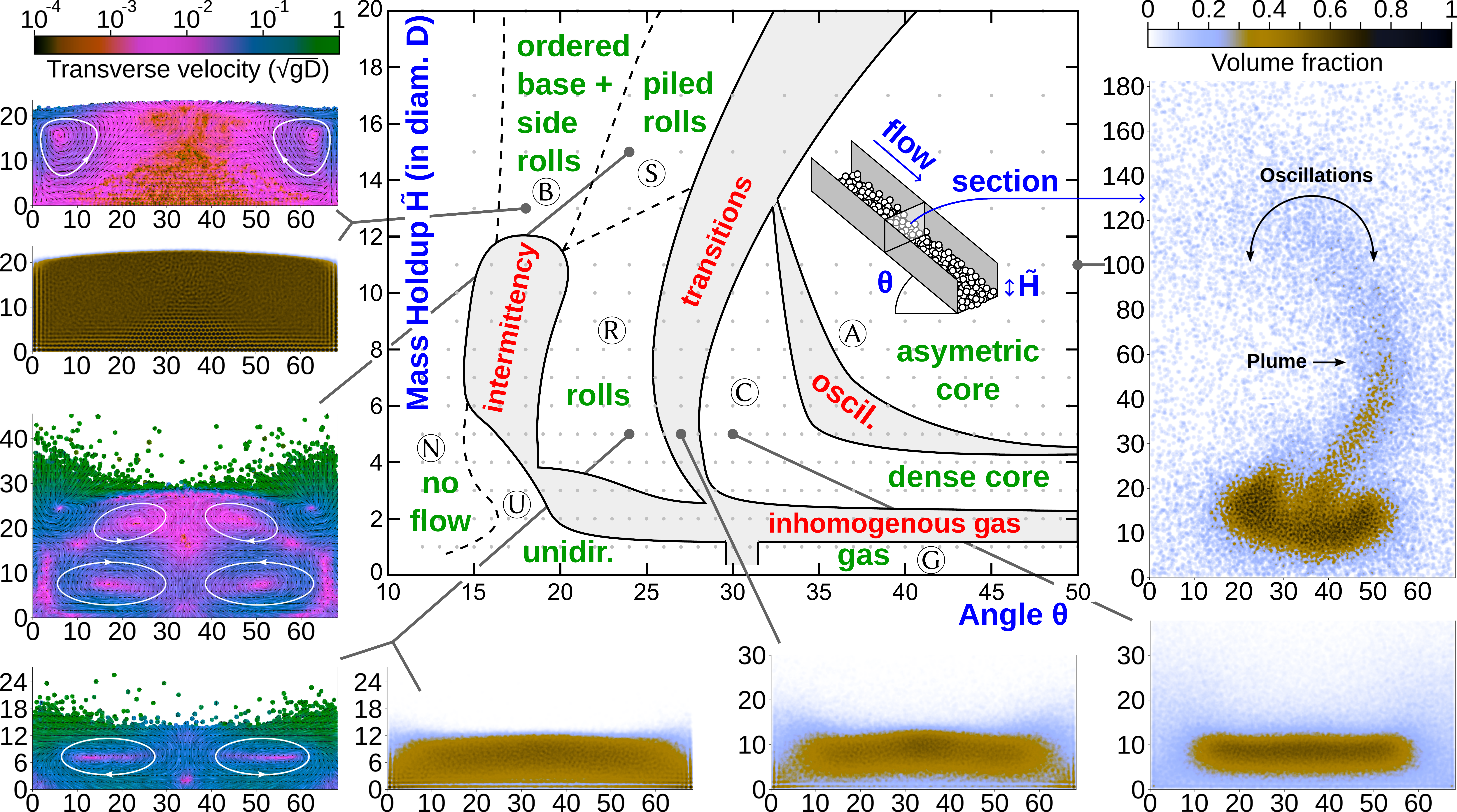}
\par\end{centering}
\caption{
Phase diagram  in the {\itshape  mass holdup - angle of inclination} space. 
\regimeU : Unidirectional flows; \regimeR : flows with Rolls; \regimeC : flows with dense Core (i.e., supported flows);
\regimeA : supported flows with Asymmetric core; \regimeS : flows with Superposed rolls;
\regimeB : flows with a Basal ordered layer toped by rolls.
The 385 gray dots are the sampling points in the phase space where we
performed a simulation (also visible in Fig.~\ref{fig:wallcontrib}). The phase diagram is supplemented with 2D maps representing the velocity in the transverse direction (left panels) and the particle volume fraction (middle and right panels). These data are averaged over the periodic direction $X$ and over $500\sqrt{D/g}$ time units, which is much larger than a typical oscillation in regime~\regimeA. The snapshot on the right is thus averaged only in $X$ and taken at a single time $t=2000\sqrt{D/g}$. The transition regions in the phase diagram are conservatively defined from the structure of the flow (e.g. bottom right middle snapshot) as well as from the evolution of the kinetic energy over time (e.g. oscillations).}
\label{fig:PhaseDiagram} 
\end{figure*}

-- Regime \regimeU~corresponds to classical Unidirectional dense flows.

-- Regime \regimeR~corresponds to flows with Rolls
previously reported in experimental and numerical
works \citep{Forterre_Pouliquen_2002,Borzsonyi_etal_2009,Brodu_etal_2013}.

-- Regime \regimeC~stands for the supported regime described in section \ref{sec:supported_regimes}.

-- Regime \regimeS~corresponds to the Superposed rolls and appears
at larger mass holdups $\tilde{H}$ than regime \regimeR. An example
is shown in the snapshots of Fig.~\ref{fig:PhaseDiagram}.

-- Regime \regimeB~is characterized by the presence of a Basal layered
structure. The observed order (see snapshots of Fig.~\ref{fig:PhaseDiagram})
is dynamically maintained by collisions and cage effects. The layers
are sheared and not static. Rolls are present in the disordered zone
on the top of the basal layers and are localized close to the lateral
walls.

All these regimes have distributions of mass and of velocities which are symmetric relatively to the central vertical plane: $Y=W/2$. At large inclination angles, this symmetry is broken as the mass holdup increases.

-- Regime \regimeA~denotes the supported regime with asymmetric core. 
The dense core swings back and forth from left to right. 
For larger $\tilde{H}$ and $\theta$, a plume eventually forms on top of the core, as shown in
the snapshots of Fig.~\ref{fig:PhaseDiagram}. The symmetry is recovered when averaging over one oscillation cycle, and this average is stationnary.

These different flow regimes open many perspectives to test the relevance
of granular rheological models. For example, our results
may be interpreted in the framework of the second-order fluid
model proposed in \citep{McElwaine_etal_2012} which predicts that shallow
flows develop curved surface, as seen for regimes \regimeB~and \regimeC.\\

\section{Scaling behavior}
\begin{figure}
\begin{center}
\includegraphics[width=0.4 \columnwidth]{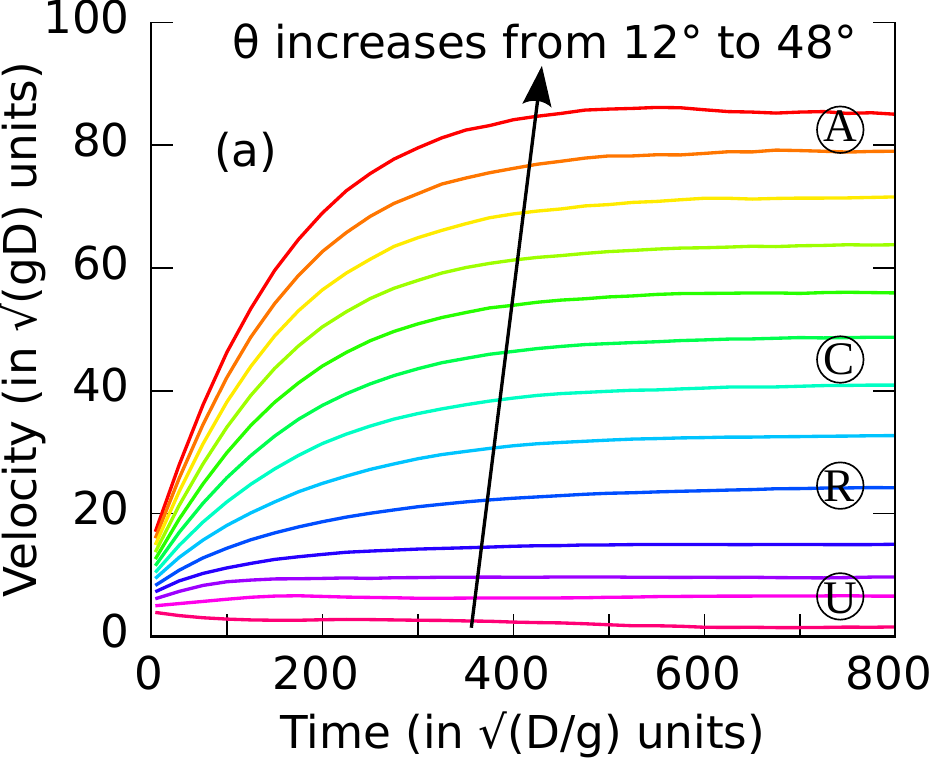}
\includegraphics[width=0.45 \columnwidth]{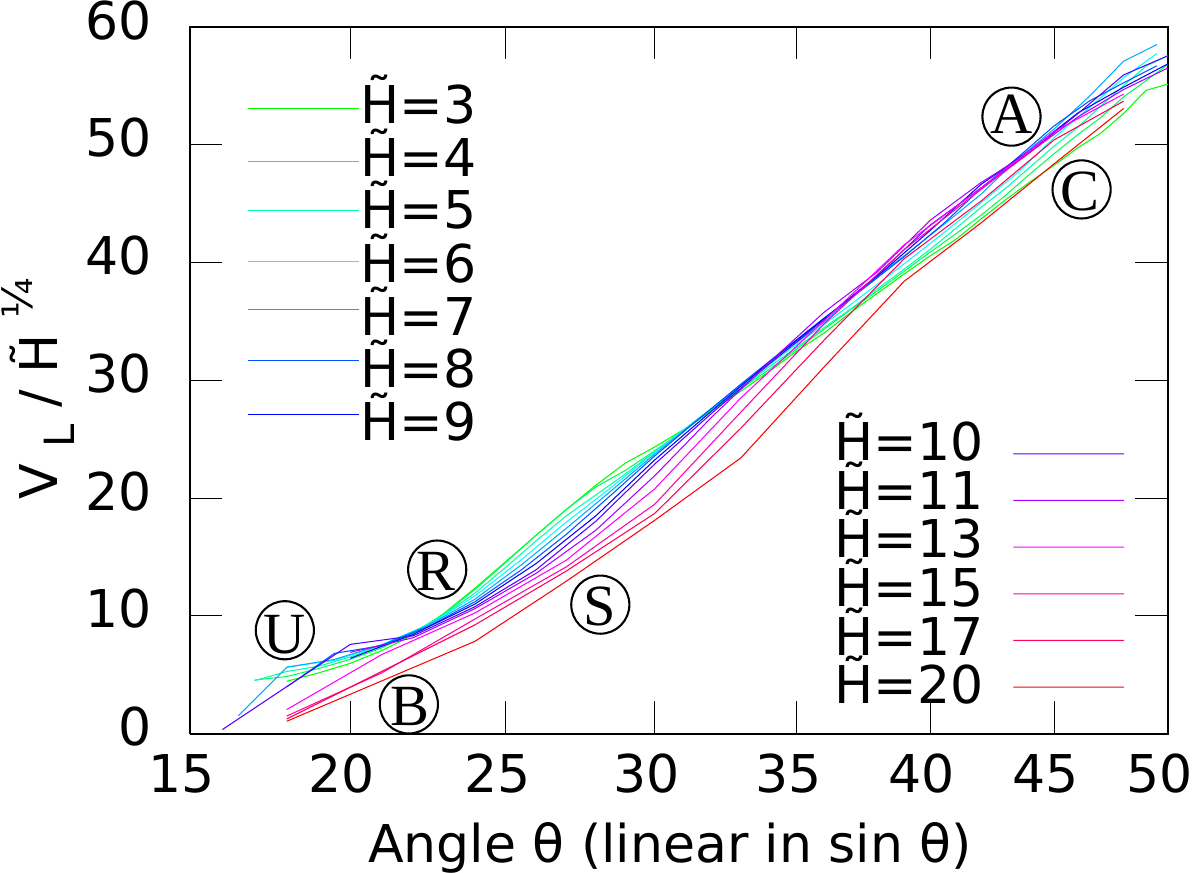} 
\end{center}
\caption{\label{fig:Velocity}
(a) Typical temporal
evolution of the mean flow velocity, for $\tilde{H}=5$ and $\theta=12$\textdegree{}
to $\theta=48$\textdegree{} every $3$\textdegree{}. All the flows reach a steady
state via an exponential saturation. We ran the simulations up to $2000\sqrt{D/g}$ time units. 
(b) Rescaled steady state velocity $V_L/\tilde{H}^{1/4}$ as a function of $\sin \theta$ for various mass hold-up.
The collapse is remarkable given the wide diversity of regimes. 
The scaling law simply reads: $V_{L}/\tilde{H}^{1/4}\approx A\sin\theta+B$, with
$A\approx122$ and $B\approx-37$. These parameters may or not depend on the channel width W, the coefficients of friction and elasticity, etc.
} 
\end{figure}

Although these flow regimes exhibit marked difference in terms
of structural organization, they surprisingly show common features. 
First, the transient regime necessary to reach the steady state
is well described by a simple exponential saturation for any value
of the inclination angle and mass hold-up:
$V(t)=V_{L}-\left(V_{L}-V_{0}\right)\exp\left(-t/\tau\right)$,
where $V(t)$ is the average streamwise flow velocity at time $t$,
$V_{0}$ the initial flow velocity,
and $V_{L}$ the limit velocity (see Fig.~\ref{fig:Velocity}a).
The characteristic time $\tau$ is an increasing function of 
the mass hold-up and has a non-monotic variation with the inclination angle (see Supplementary Figure~4a). 
This exponential velocity saturation, observed in all the regimes, suggests that the flow experiences a viscous-like drag force proportional to the velocity (see Supplementary Figure~4b).

Second, we identify a simple dependency
of the limit velocity on the mass hold-up and inclination angle. At any fixed angle, the velocity follows a power law $\tilde{H}^\alpha$, with an exponent $\alpha$ weakly dependent of $\theta$ but close to 0.25. Figure~\ref{fig:Velocity}b,
reports the limit velocity $V_{L}$ rescaled by $\tilde{H}^{1/4}$
versus the inclination angle $\theta$ for various mass
holdups.  The observed collapse is remarkable given the large diversity of the flow regimes. In steady state, the mass flow rate is simply
given by $Q=V_{L}\tilde{H}$ such that $Q\propto\tilde{H}^{5/4}$.
In the configuration of regime~\regimeU, \citep{Louge_Keast_2001} have 
experimentally measured an exponent $Q\propto\tilde{H}^{3/2}$, which is within
the range of exponents shown in Figure~\ref{fig:Velocity}b.\\

\section{Conclusion}
Using a simple flow configuration with flat lateral and basal boundaries,
we have discovered, by increasing the inclination angle and mass holdup, steady and fully developed regimes which
present non-trivial features including heterogeneous volume fraction, secondary flows,
symmetry breaking and dynamically maintained order.
Despite the diversity of the features
of these states, we have highlighted that the mass flow rate obeys
a scaling law in terms of $\tilde{H}$. 
Explaining these regularities is a challenging issue, as they suggest
a unified underlying model.  

A crucial question is to which extent these regimes and their features
are specific to the material parameters and the confined geometry 
we have considered. Additional simulations, where we have varied the
material parameters (friction and restitution coefficient) and
the basal conditions (flat or bumpy), lead to similar regimes 
as long as grain/wall friction prevails on grain/grain friction. 
Side walls play of course an important role regarding the friction and allow 
SFD flows for any value of the chute inclination. Without side
walls, at least for flat bases, flows at large angle would not be steady but accelerated.
Despite of this, analog flow regimes appear but as transient state (see Supplementary Figure 5).

These results provide a unique set of very complex granular flow regimes for testing theoretical and rheological models. These regimes surprisingly appear in a configuration, the inclined channel, that was previously considered boring and well-studied. We have however only explored a small portion of the full high-dimensional phase diagram consisting of the variations of all influencial parameters. It it thus very likely that more regimes exist, especially in the wide range of conditions found in nature and in industry. We hope that our study will encourage such investigations of granular flows, in particular for wider channels and higher mass holdup values.\\

This work was partly financed by the RISC-E RTR and R\'egion Bretagne
(CREATE Sampleo grant). We thank Michel Louge, Jim McElwaine, Jim Jenkins, Anne Mangeney and Olivier Roche for helpful discussions and comments on our work.

\appendix
\section*{Appendix. Derivation of the position of the center of mass of the flow as a function of the inclination angle\label{sec:balance}}

The linear variation of the center of mass $C_{M}$ of the flow as
a function of the tangent of the inclination angle can be interpreted
by considering a simple force balance. In a stationary regime, we
have : $Mg\sin\theta=2\mu_{w}L\int_{0}^{\infty}N_{w}(z)dz+\mu_{b}N_{b}WL$,
with $N_{w}(z)$ and $N_{b}$ the normal stress at side-walls and
at the base respectively. The first term of the right-hand side is
the friction of the wall on the flow, and the second one is the friction
at the base. Both $\mu_{w}$ and $\mu_{b}$ saturate in the supported
regimes (see Fig.~2a), so they can be considered as constant in a
first order approximation. Using the fact that $Mg\cos\theta=N_{b}WL$,
we end up with: $2\int_{0}^{\infty}N_{w}(z)dz/N_{b}=\frac{W}{\mu_{w}}\left(\tan\theta-\mu_{b}\right)$.
The term of the right-hand side represents a characteristic height
of the flow, denoted later on by $H_{p}$ (if the pressure was purely
hydrostatic, $H_{p}$ would correspond exactly to the height of the
flow). The derived relation expresses the fact that the characteristic
height of the flow $H_{p}$ varies linearly with $\tan\theta$. This
explains why the position $C_{M}$ of the center of mass, which is
of the order of the half characteristic height of the flow, evolves linearly with
$\tan\theta$. Remarkably, we find here a slope $a\approx W/\left(2\mu_{m}\right)$
in agreement with the hypothesis that $C_{M}\thickapprox H_{p}/2$.
We also find that the "atmospheric height" $H_{C}$ increases
linearly with $\tan\theta$ in the supported regimes. The slope here
is about two third of $W/\mu_{w}$, corresponding to $H_{c}\thickapprox3H_{p}/2$.

\bibliographystyle{jfm}

\bibliography{jfm_slows_v10}

\end{document}